\documentclass[journal=acsnano,manuscript=article]{achemso}

\usepackage[version=3]{mhchem} 
\usepackage{xcolor}

\author{Olivier No\"el}
\email{olivier.noel@univ-lemans.fr}
\affiliation{%
IMMM, UMR CNRS 6283, Le Mans Universit\'e, Av. O. Messiaen, 72085 cedex 09, Le Mans, France
}%
\author{ Pierre-Emmanuel Mazeran}
\affiliation{
Sorbonne universit\'es, Université de Technologie de Compi\`egne, UMR CNRS 7337, Roberval, Centre de recherche de Royallieu – CS 60319–60203 Compi\`egne cedex, France
}%
\author{Igor Stankovi\'{c}}
\email{igor.stankovic@ipb.ac.rs}
\affiliation{
Scientific Computing Laboratory, Center for the Study of Complex Systems,
Institute of Physics Belgrade, University of Belgrade,
11080 Belgrade,
Serbia
}

\title[Nature of dynamic friction in a humid hydrophobic nano-contact]
  {Nature of dynamic friction in a humid hydrophobic nano-contact}

\abbreviations{IR,NMR,UV}

\begin{document}

\begin{tocentry}

\includegraphics[width=8.2cm]{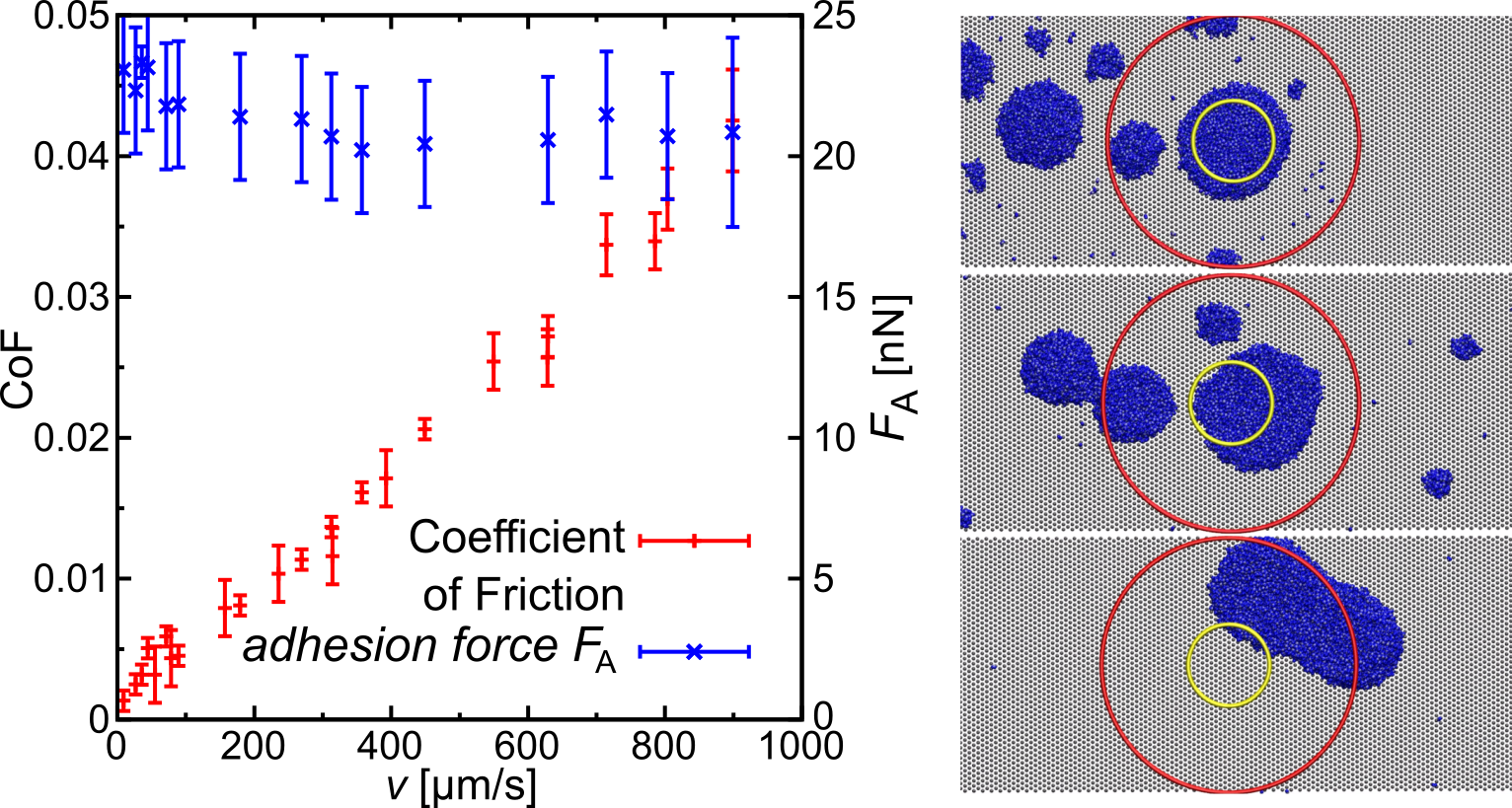}

\end{tocentry}

\begin{abstract}
The physics of dynamic friction on water molecule contaminated surfaces is still poorly understood. In line with the growing interest in hydrophobic contact for industrial applications, this paper focuses on friction mechanisms in such interfaces. As a commonly used material, contact with graphite is considered in a twin-fold approach based on experimental investigations using the circular mode atomic force microscopy technique combined with molecular dynamic simulations. We demonstrate that an intuitive paradigm, which asserts that water molecules are squeezed out of a hydrophobic contact, should be refined. As a consequence, we introduce a mechanism considering a droplet produced within the sliding nano-contact by the accumulation of water adsorbed on the substrate. Then we show that a full slip regime of the droplet sliding on the hydrophobic substrate explains the experimental tribological behavior.

\textbf{Keywords:} dynamic friction, graphite, circular mode, atomic-force microscope, molecular dynamics

\end{abstract}

\section{Introduction}
The investigation of water-mediated forces involved in tribological processes in hydrophobic nano-contacts arouses an undeniable interest both for technological applications and from a fundamental point of understanding of the nature of these interactions~\cite{2008-Jinesh,2016-Ye,2016-Vilhena,2016-Rhee,2018-Arif}. For instance, dry lubricants or coatings developed in response to technological and ecological issues, and driving the demand for chemically stable, high-performance or efficient in dusty environment systems, are frequently hydrophobic. Among sectors of the economy concerned by such developments are automotive, aerospace, steel, oil and gas, mining and mineral processing, energy and power industries. In daily use, the role of humidity in friction becomes critical~\cite{2016-Rhee,2018-Arif,2018-Carpick,2012-Nguessan}. As another illustration, hydrophobic nano-contacts are an active part of human interface systems involving the epidermic friction that increases with humidity of the finger pad\cite{2011-Pasumarty}.

In this paper, we focus on graphite, a commonly used hydrophobic material for industrial applications. It is representative of a class of hydrophobic materials, such as polytetrafluoroethylene (PTFE), talc, and hexagonal boron nitrate. We consider friction between the highly oriented pyrolitic graphite (HOPG) and a nitride silicon atomic force microscopy (AFM) probe. The HOPG is atomically flat with very few atomic steps when freshly cleaved. After exposure to air for roughly thirty minutes, the HOPG behaves like a hydrophobic substrate, i.e., the contact angle with water $80^o-95^o$, cf. Ref.~\cite{Li2013}. 
In such materials, both friction and wear are strongly influenced by humidity, and several mechanisms are proposed. Adhesion and friction forces are substantially affected as water condenses on surfaces and into nano spaces altering the contact area at the nanoscale. The growth of water condensate within the contact hinges on a kinetic process~\cite{2018-Vitorino,2005-Riedo}. The condensation mechanism occurs rapidly considering the experimental time scales, while the evaporation remains quite slow~\cite{2007-Feiler}. Nevertheless, condensation alone cannot explain the overall tribological mechanisms in hydrophobic contacts, characterized by weaker interactions with water molecules. 
Although it was demonstrated that the pinning–depinning molecular process induces dynamic shear forces in hydrophilic systems~\cite{2015-Lee}, one should expect slip in non-wetting or hydrophobic systems~\cite{2011-Ho,1989-Thompson,2001-baudry}. 
Only a few experimental and theoretical studies of friction in the presence of water for hydrophobic systems are available ~\cite{2010-Teschke,2016-Rhee,2016-Vilhena,2016-Ye,2018-Carpick,2018-Arif}. In particular, Hasz et al.\cite{2018-Carpick} reported a non-monotonic friction force trend with increasing humidity. A change in nature of the contact, occurring at a maximum of the friction force may explain such a behavior\cite{2007-Feiler}. Even at low relative humidity, friction results from the water capillary bridge driven by asperities sliding on the substrate: thus, the friction force is correlated to the amount of water. Whereas, with very high relative humidity, it is related to the asperities sliding on quasi-continuous water layers~\cite{2018-Carpick}. \textcolor{black}{The studies regarding the friction behavior on strongly adsorbed contaminants such as water report a highly nonlinear dependence with the normal load, i.e., a non Amontons’ law~\cite{2008-Jinesh,2010-Teschke,2010-Khan}. Such a finding is associated with the way the contact is formed: the solid slides on one or more water layers while the phase state of water and number of layers depends on the rate of loading~\cite{2010-Khan}. The phase transition of water layers into ice under both high compression and high compression rates is also extensively studied\cite{2006-Giovambattista,2018-Ouyang,2010-Khan, SHIMIZU201887}.}
Also, regarding the competition between water-water and water-substrate interactions, the molecular level interactions must be considered to fully describe the mechanism by which water comes into contact and the associated kinetics. The atomic force microscope (AFM) is a powerful tool enabling the investigation of a mono-asperity contact between two solids and measuring friction and adhesive forces with a high resolution at the nanoscale.

In following experimental dynamic friction force is obtained on hydrophobic highly oriented pyrolitic graphite (HOPG) surfaces in air, using the circular mode atomic force microscopy (CM-AFM). The CM-AFM permits accessing high sliding speeds, up to 0.8~mm/s, which is several decades larger than standard AFM studies and investigating tribology in stationary conditions\cite{Noel-2011,2012-Noel}. The experimental results are confronted with molecular dynamic (MD) simulations to highlight relevant friction mechanisms. The substrate surface is never completely dry due to the attractive van der Waals interactions between water molecules and the solid material. We assume that water adsorbed on substrate hydrophobic substrate plays a role in the formation of the capillary bridge. 



\section{Results}

\begin{figure}[ht!]
\includegraphics[width=8cm]{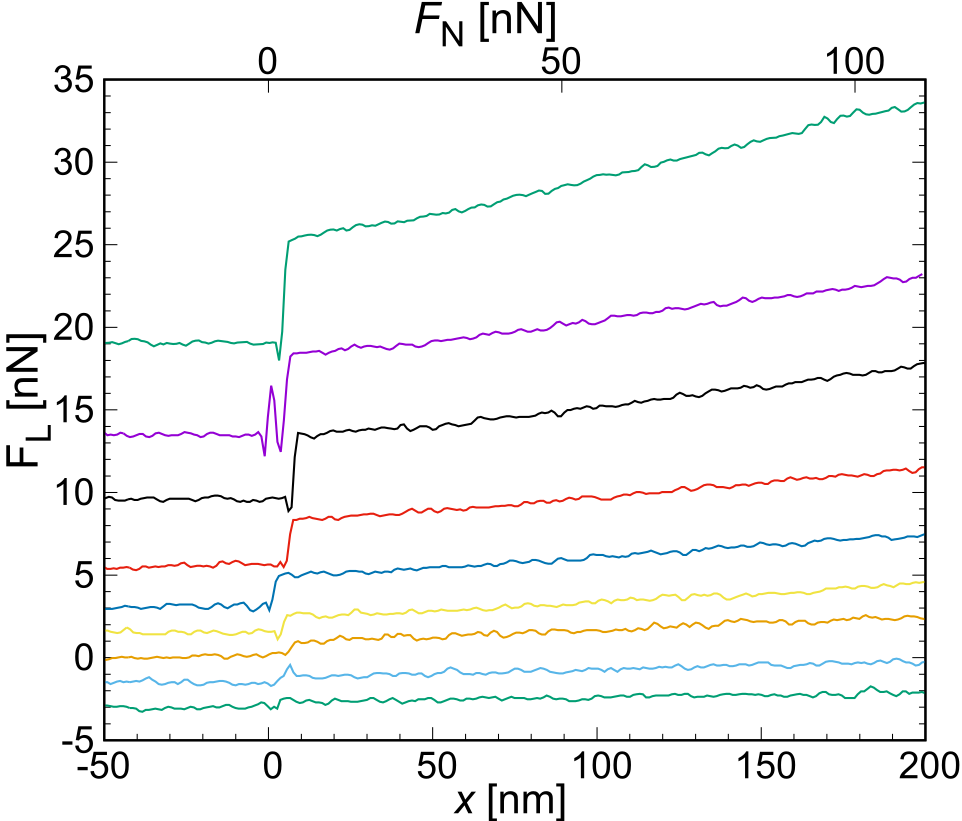}
\caption{AFM lateral force spectra or friction laws (friction force {\it vs.} normal load $F_{\rm N}$) obtained during the approach, with a 100~nm radius silicon nitride probe at relative humidity {\it RH}=$38\%$, on a HOPG hydrophobic surface with roughness $R_{\rm a} = 0.05$~nm. The results are shown for sliding velocities $v=25,50,100,200,300,400,500,700$ and $800$~\textmu m/s, respectively from bottom to top. For clarity, the curves are shifted along the y-axis. Although values of lateral force along the y-axis are arbitrary, the scale allows a relevant assessment of the lateral force values between the curves. Out of contact, {\it i.e.}, for $x<0$, the lateral force is zero (i.e, there is no contact between the probe and the substrate). The initial lateral force when contact occurs is referred to as $F_{\rm L, 0}$.}
  \label{fig:data}
\end{figure}
\begin{figure}[ht!]
\includegraphics[width=16cm]{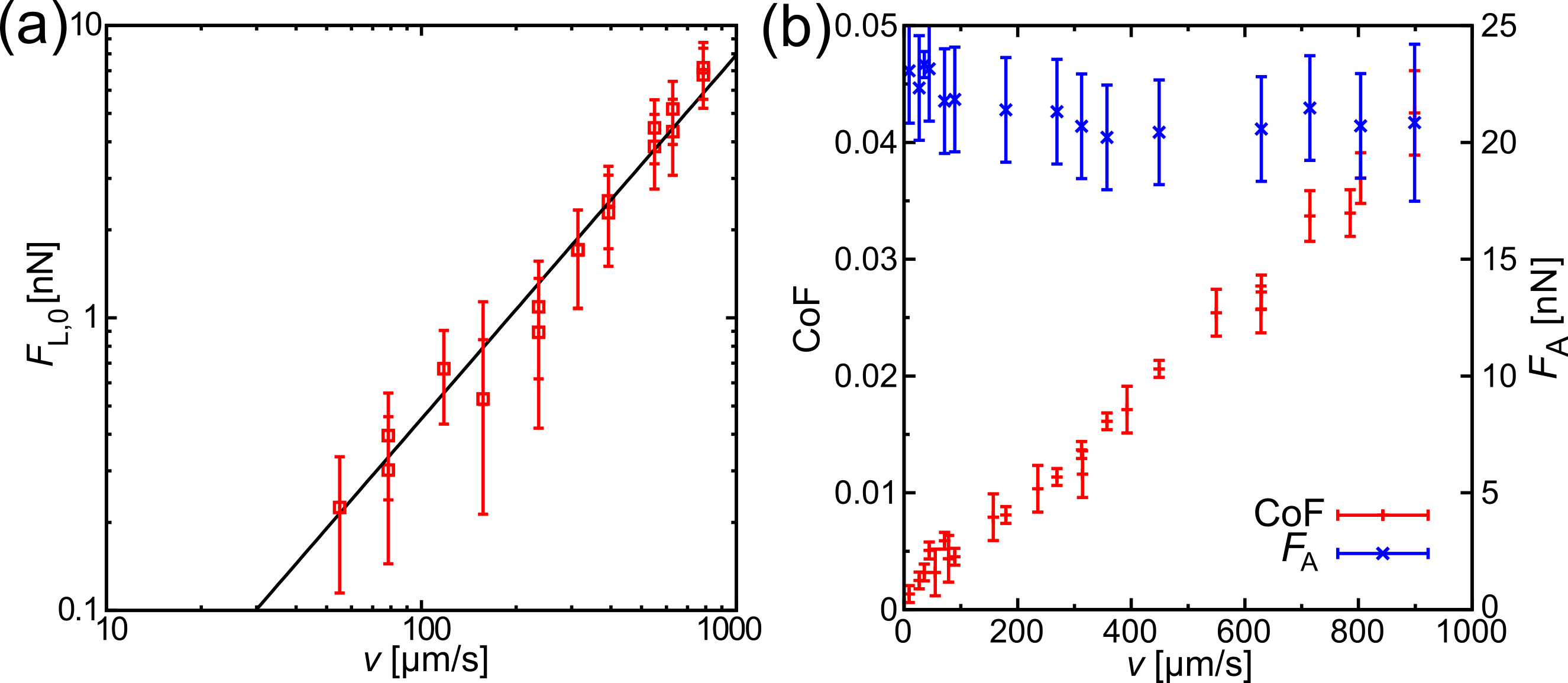}
\caption{(a) Initial experimental lateral force $F_{\rm L,0}$ (main plot) dependence on probe velocity $v$ in case of hydrophobic surfaces. The scale of the axis is logarithmic. \textcolor{black}{Linear fits through these points highlight a scaling law with the sliding velocity $F_{\rm L,0} = \eta (v/{\rm v}_0)^\alpha$ with a corresponding damping coefficient $\eta=1.5\pm0.4$~pN, $\alpha=1.24\pm0.1$ and ${\rm v}_0=1$~\textmu m/s.} (b) Variation of both the adhesion force $F_{\rm A}$ and the coefficient of friction (CoF) against the sliding velocity on HOPG. Experiments were carried out with a silicon nitride probe ($R\approx100$~nm) at a relative humidity {\it RH}=38$\%$, on a freshly cleaved HOPG hydrophobic surface with roughness $R_{\rm a} = 0.05$~nm.}
  \label{fig:cof}
\end{figure}

Friction mechanisms involving a hydrophobic contact in a given humid environment ({\it RH}=$38\%$) have been investigated with the CM-AFM. Figure~\ref{fig:data} reports typical friction force spectra or laws (lateral force {\it vs.} normal load) for different sliding velocities ranging from 50~\textmu m/s up to 0.8~mm/s during the approach of the probe to the surface. \textcolor{black}{Actually, similar friction force curve behaviors are obtained both during the approach and retraction of the probe. The adhesion introduces only an additional normal load (see Figure S2 in SI). In the case of lubrication where no solid-solid contact occurs, friction force dependence is expressed by the Derjaguin form\cite{2015-Eder}, $F_{\rm L}=F_{L, 0}+\mu F_{\rm N}$; i.e., a non-vanishing offset $F_{\rm L,0}$ complements the Amontons-Coulomb term $\mu F_{\rm N}$.} The set of spectra clearly shows a {\it jump into} contact ($F_{\rm L,0}$) in the lateral force as a contact becomes effective. \textcolor{black}{The experimental data follow a power-law which can be expressed as: $F_{\rm L,0} = \eta (v/{\rm v}_0)^\alpha$ with a corresponding damping coefficient $\eta=1.5\pm0.4$~pN, $\alpha=1.24\pm0.1$ and ${\rm v}_0=1$~\textmu m/s, see Figure~\ref{fig:cof}(a). Physically, the damping parameter $\eta$ represents the energy dissipated by the moving water bridge.} This highlights the presence of water within the contact and accounts for a significant change in its nature. Indeed, considering a dry contact between non-viscoelastic solids, friction is weakly (or logarithmically) dependent on the sliding velocity~\cite{Ptak2019}. Moreover, if the probe is partially hydrophilic (i.e., by considering a contact angle with water of 70$^o$ as the one we should expect for a clean nitride silicon substrate), similar power-law behavior is observed by simulation, and the mechanisms described are not different. Indeed, the material exhibiting weaker interactions with water determines the friction behavior of the overall system.

\begin{figure}[ht!]
\includegraphics[width=8cm]{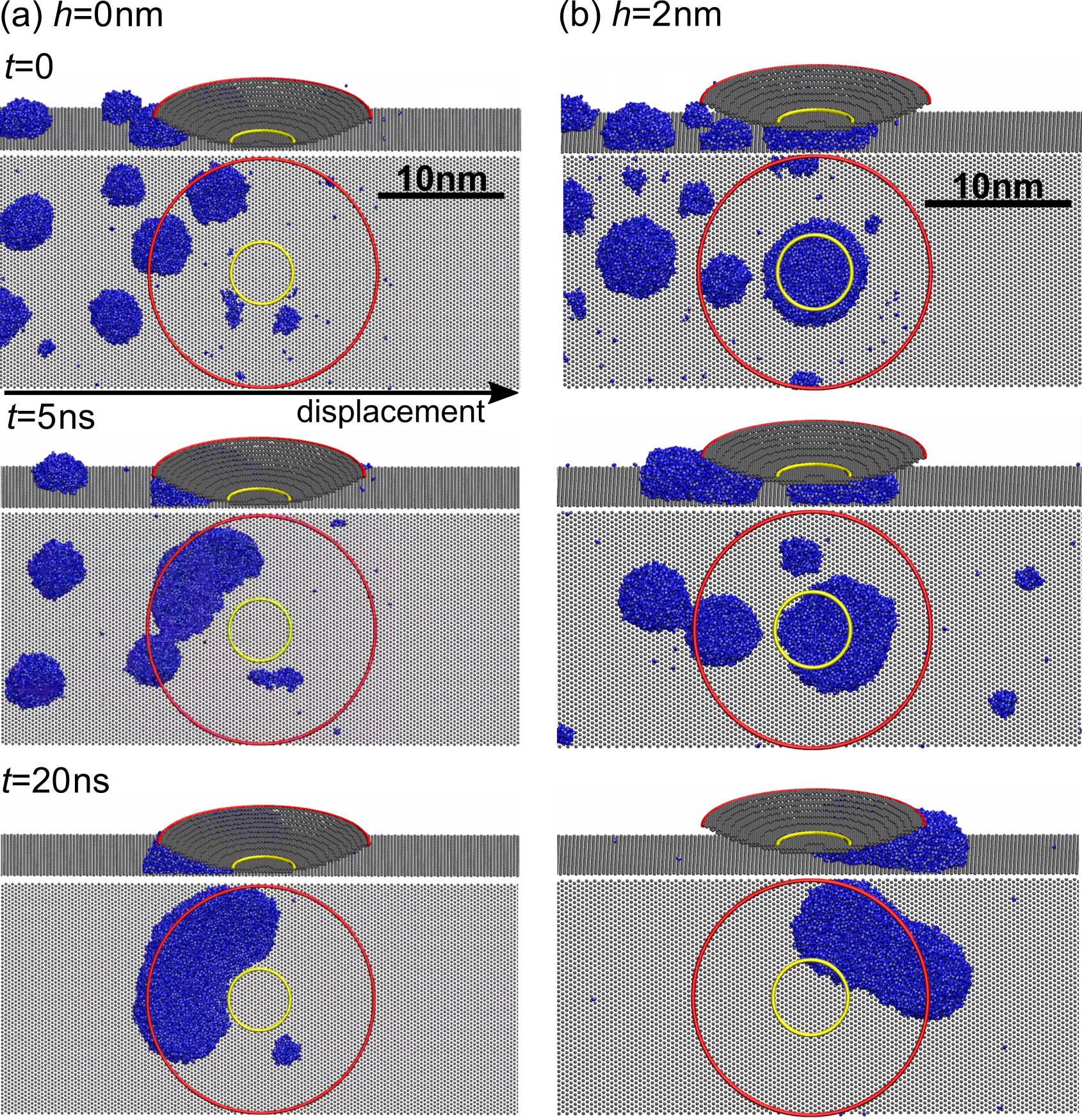}
\caption{Snapshots of water configuration for two different probe-substrate separations (a) $h=0$~nm (left panels) and (b) $2$~nm (right panels) between the probe and the sample. For convenience, in these simulations, the substrate moves with velocity of $v_s=1$~m/s from left to right of the panels as indicated by the arrow. The volume of water is constant throughout the simulation.}
  \label{fig:snaps}
\end{figure}

The dependence of the onset lateral force with the sliding velocity can stem either from water-substrate interactions or viscosity of water inside capillary bridge. The viscous hydrodynamic damping properties inside the water nanodroplet are a result of stress fluctuations due to interactions between water molecules, according to the Green-Kubo mechanism~\cite{green,kubo}. The viscous drag force at a 1~\textmu m/s sliding velocity with a nanodroplet radius in the range of 1~nm-10~nm has been calculated considering the usual assumption of a no-slip boundary condition while accounting for the bulk viscosity of water\cite{2015-Kim,2015-Lee}. It was estimated to be 10$^6$-10$^7$ times smaller than the maximal experimental lateral force value (about 10~nN: see Figure~\ref{fig:data}). Therefore, such viscous effects cannot explain the experimental data: thus, water-substrate interactions must be scrutinized. First, the power-law dependence of the onset lateral force against the sliding velocity emphasizes the irrelevance of considering a thermally activated process of pinning of water molecules on the substrate surface atoms ~\cite{2001-baudry, 1997-Thompson, 1989-Thompson}. This contradicts the classical paradigm accounting for surface intercalation or pinning as a source of lateral forces through capillary water bridges both in hydrophilic and hydrophobic systems~\cite{2015-Kim,2015-Lee}. Indeed, the intercalation with two-dimensional materials is associated with the adsorption of water molecules on the surface yielding a logarithmic (weak) dependence on the sliding velocity of the lateral force. The power-law behavior (observed as almost linear) may arise from the van der Waals interactions between water and the substrate. These interactions are insufficient to cause strong bindings between water molecules and the substrate. Therefore, the lateral force stems from collective interactions entailing a specific slipping resistance within the water-substrate contact. 

In Figure~\ref{fig:cof}(b), CoF values are determined by the slope of the $F_{\rm L}$ {\it vs.} $F_{\rm N}$ curve for different sliding velocities (from AFM lateral force spectra in Figure~\ref{fig:data}). The CoF linearly increases with the sliding velocity. The system exhibits a low CoF, i.e., superlubric regime CoF$<$0.01 at low velocities ($v <200$\textmu m/s). The observation that the effective coefficients of friction (CoF) depend on the sliding velocity indicates that water droplet resistance to slip is responsible for the dependence of the lateral force on the sliding velocity.
 
In the following, molecular dynamics simulations have been performed to get a better understanding of the mechanisms involving the water-substrate interactions. Figure~\ref{fig:snaps} shows the results of molecular dynamics simulations for a hydrophobic probe-sample contact. These simulations are initiated with a homogeneously distributed water on the surface and the water adsorbs at the surface through van der Waals forces (described by the Lennard-Jones potential) upon the hydrophobic substrate.

At the molecular level, cohesive forces between water molecules prevail over their attraction to the hydrophobic substrate. As a result, water molecules gather to form water droplets adsorbed either on the free substrate or inside the probe-substrate gap. Water droplets are driven by the motion of the sliding probe and aggregate with the water clusters adsorbed between the probe and substrate entailing a growth mechanism (see Figure~\ref{fig:snaps}). When the water droplet, gathered below the probe, outgrows the height of the gap, it will be squeezed out from the area of the closest approach. This observation is in agreement with previous simulations involving hydrophobic surfaces~\cite{2016-Ye,2016-Vilhena}. The aggregation of water into clusters and smaller droplets captured or driven by the probe during sliding conveys the principal process of water droplet formation and growth inside the contact. Moreover, one should notice that depending on the probe-substrate distance, the shape of the accumulated water droplet may be quite different. 
At the beginning of the process, a profile of small droplets (relative to gap size) adsorbed on the plane substrate and dragged inside the gap between the probe and the substrate is circular (see Figure~\ref{fig:snaps} at $t=$0~ns). Then, the droplets become elongated as they grow. The substrate pulls full-grown water droplet to the extremity of the probe in the movement direction (see Figure~\ref{fig:snaps}(b) at $t=$20~ns). The outlined scenario shows that the condensation process of water molecules is not a mandatory condition to form a capillary bridge in the contact. Here, the mechanism of water accumulation by moving probe is derived from simulations carried out with high sliding velocities, i.e., 1~m/s, compared to the experimental values remaining typically below 1~mm/s. That also implies a calculated time of displacement averaging 20~ns. During the formation of the water capillary bridge, simulations show that water molecules leave the droplets before re-adsorption either on the substrate or back onto the other clusters of water. Furthermore, the measured lateral forces may also be affected by evaporation mechanisms. However, with currently available computational resources, we cannot bridge three orders of magnitude in the computational time necessary to reach the steady-state for a comprehensive assessment of the influence of the evaporation process on the water capillary bridge organization. \textcolor{black}{ We should also note a difference with the systems completely covered with a layer of water~\cite{2016-Vilhena,2008-Jinesh}. Indeed, in the outlined scenario, we do not expect the formation of thin water layers. The surface forces squeeze the accumulated water out of the contact, while simultaneously the water droplet in the gap and attached to the surface is pulled out by the motion of substrate (cf. Figure~\ref{fig:snaps}).}

\begin{figure}
\includegraphics[width=16cm]{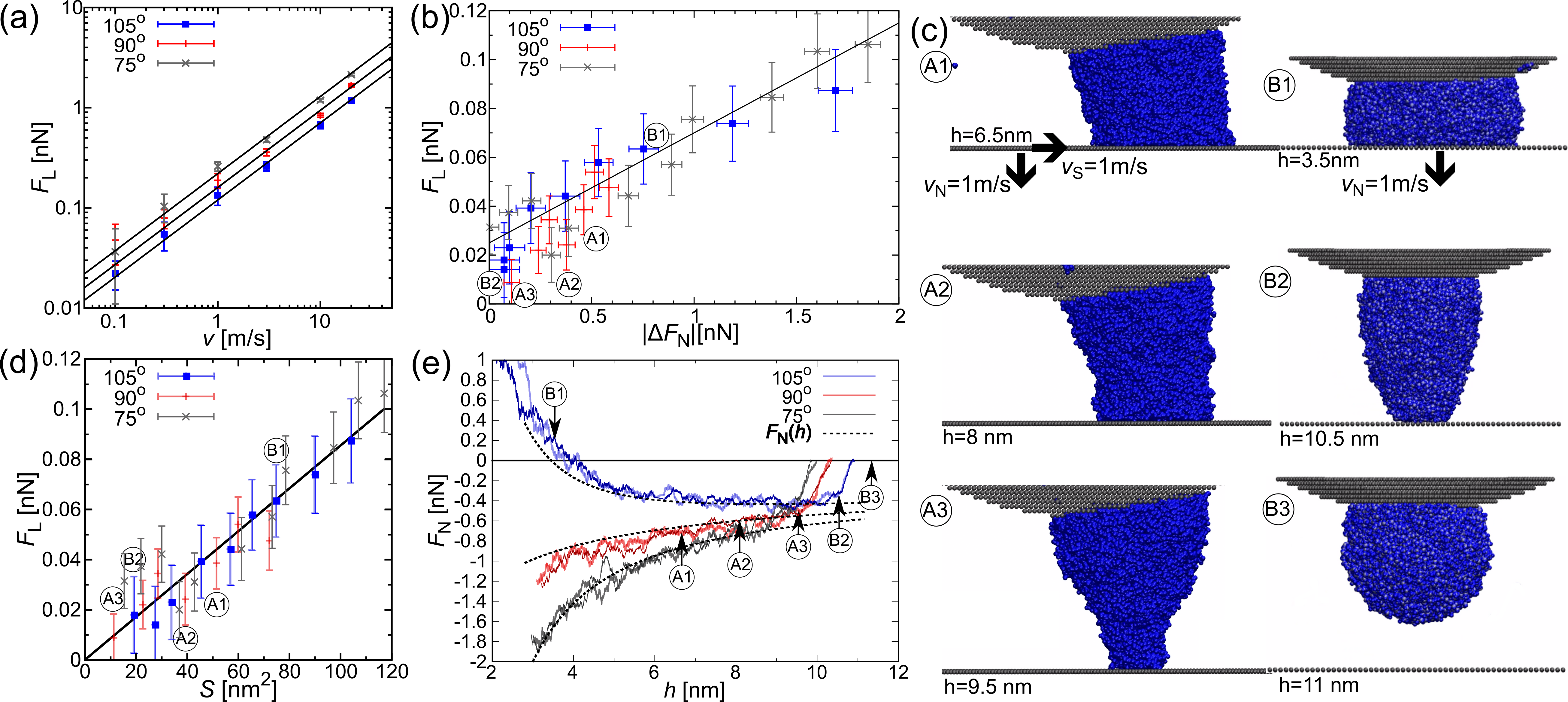}
\caption{(a) The calculated lateral force $F_{\rm L}$ against velocity $v$.  The results are conveniently described with a power law $\alpha=0.77\pm0.03$. (b) Lateral force $F_{\rm L,0}$ spectra vs. normal force $F_{\rm N}(h)$ for three contact angles, $\theta=75^{\rm o}, 90^{\rm o}$, and $105^{\rm o}$ at sliding velocity $v_{\rm S}=1$~m/s. (c) The left-side panels show configuration snapshots of the water bridge for different probe-substrate distances $h=6.5, 8,$ and $9.5$~nm respectively denoted A1, A2, and A3 for water-substrate contact angle $\theta=90^{\rm o}$ and sliding velocity $v_{\rm S}=1$~m/s. The right-side panels show water bridge for water-substrate contact angle $\theta=105^{\rm o}$ and laterally static probe ($v_{\rm S}=0$~m/s). The snapshots are shown for different probe-substrate distances $h=3.5, 10.5$ and $11$~nm, respectively, denoted B1, B2, and B3. The configurations are also reported on the curves. (d) Lateral force $F_{\rm L,0}$ dependence on water contact surface area $S$. The points are obtained for three surfaces for sliding velocity $v_{\rm S}=1$~m/s. Linear fit through these points highlights a linear dependence on surface area $S$. (d) Results of simulations (full lines) of the retraction of the AFM probe with velocity $v_{\rm N}=1m/s$ are compared with $F_{\rm N}(h)$ calculated with the analytic model (dashed line). The simulations were performed for laterally static probe $v_{\rm S}=0$~m/s (thick) and with sliding velocity $v_{\rm S}=1$~m/s (thin line). All simulations are performed with a 340~nm$^3$ water droplet and for different contact angles $\theta=75^{\rm o},90^{\rm o}$, and $105^{\rm o}$.}
\label{fig:philicphobic}
\end{figure}

The Figure~\ref{fig:philicphobic}(a) reports the evolution of the lateral force $F_{\rm L}$ against the sliding velocity values calculated by molecular dynamics simulations for 340~nm$^3$ nanodroplet. The computed lateral force values follow a power-law similar to the measured ones considering both contact angles (as with simulations: $\theta=$75$^{\rm o}$, 90$^{\rm o}$, and 105$^{\rm o}$), with a calculated quantity $\alpha=0.77\pm0.03$ instead of $1.24$ for the experimental data. A decrease of the lateral force with $\theta$ is also noticed. Regarding simulations, the contact surface does not depend on the sliding velocity. Thus, the observed difference of exponents between both experimental and simulated trends suggests that the actual contact area may slightly increase with the sliding velocity. The water droplet volume can be determined by comparing the simulations and experiments according to Figure~\ref{fig:cof}(a). However, such a comparison overestimates this volume. Indeed, for a given sliding velocity, the droplet should have a contact surface averaging $S\approx6000$~nm$^2$ which is a quite high value considering the probe diameter (0.2~\textmu m). Hence, other mechanisms should be examined to explain such a discrepancy with further studies. 

In regard to the simulations, the average velocity of water molecules in contact with the substrate is equal to the average velocity of the water capillary bridge, which confirms that the full-slip mechanism occurs at the contact. A quasi-viscous dependence of the lateral force on the sliding speed arises from the water molecules' resistance to slip at the droplet-substrate contact. Besides, it is interesting to underline that the close to the linear dependence of the lateral force with the sliding velocity is confirmed for all studied contact angles ($\theta=$75$^{o}$, 90$^{o},$ and 105$^{o}$), see Figure~\ref{fig:philicphobic}(a). \textcolor{black}{Moreover, simulations performed with the probe both sliding and retracting relatively to the substrate, show that lateral forces depend linearly on the normal force and contact surface area, cf. Figure~\ref{fig:philicphobic}(b) and (d), respectively. However, Figure~\ref{fig:philicphobic}(c) shows that the lateral force increase correlates to the change in the contact area induced by the applied load.} Indeed, in Figure~\ref{fig:philicphobic}(c), configuration B1 corresponds to the compressed droplet exerting a repulsive force ($F_{\rm N}>0$) between the probe and substrate. The same droplet in configuration B2 is pulled apart jointly by the probe and substrate. The resulting normal force in the latter configuration (B2) is, therefore, adhesive ($F_{\rm N}<0$). However, the extension of the droplet results in a smaller contact surface compared to case B1. In conclusion, the water droplet on the hydrophobic surface is behaving as an incompressible system and the application of a normal force yields an elastic deformation of the droplet modifying its contact surface, cf. Figure~\ref{fig:philicphobic}(c), and in turn increasing so the lateral force. Moreover, the congruence between the simulated and experimental behaviors suggests that the probe behaves as a hydrophobic material. This assumption is confirmed by electron scanning electronic microscopy images that show the probe contaminated probably by a hydrocarbon layer (see Figure S1 in SI). To this end, by choice, the used probes were not voluntarily decontaminated before the experiments to obtain a hydrophobic/hydrophobic contact.

Previous observations raise the issue of the role of evaporation in the evolution of water droplets. We remind, that the probe is always in contact with the surface in the experiments. This contact can be direct or indirect through a water bridge. In addition, the substrate is continuously scanned underneath the probe with a high sliding velocity compared to that of the loading. The moving capillary bridge forms and grows by the accumulation of water clusters and droplets covering the substrate and encountered on its path~\cite{2012-Verdaguer}. An apt description of the water-aggregation process must be considered in two dimensions. \textcolor{black}{In other words, one could imagine “a kind of growing packman swallowing small water droplets.” Still, moderate $F_{\rm L}\propto v^{1.24}$ power-law dependence of the lateral force on velocity indicates that there are processes limiting the growth of the water bridge by accumulation. For example, as the moving capillary bridge grows, its surface increases, simultaneously as the evaporation process induces a consequent water loss and thus, limiting the growth of the water bridge.} The Figure~\ref{fig:philicphobic}(a) represents simulations obtained for a fixed amount of water in the system - not accounting for the growth of the droplet by accumulation. The simulations are also performed at high sliding velocities yielding conditions where the evaporation has no effect and viscous drag and surface adhesion limit the size of the droplet. Therefore, a $F_{\rm L}\propto v^{0.77}$ behavior could be considered as a lower limit for the lateral force dependence on the sliding velocity in our system while considering that volume of the capillary bridge does not change with the sliding velocity, cf Ref.~\cite{2015-Lee}. 



The adhesion force provides another way to estimate the quantity of water present in the contact. Simulations in Figure~\ref{fig:philicphobic}(d) focus on the retraction of the AFM probe for both a sliding and a static contact. The simulations are performed considering a constant 340~nm$^3$ water volume and at two different sliding velocities, $v=$0~m/s, and 1~m/s. Figure~\ref{fig:philicphobic}(d). The panels show snapshots of two water bridges at different probe-substrate distances, cf. Figure~\ref{fig:philicphobic}(c). One should notice that the contact angle with the water droplet thermally fluctuates. Also, the sliding probe pulls the droplet slightly out of the center (A1-A3), while for static substrate droplet is directly under the probe (B1-B3). Panels B2 and B3 show ($h=10.5$~nm and $11$~nm, respectively) depict the capillary bridge configurations just before and just after the jump-off of the droplet (i.e., point when the capillary bridge departs from the substrate). After the jump-off, the water droplet remains linked to the probe. The adhesion force, given by the jump-off represented by the two points B2 and B3 on the curve, is almost independent of the sliding velocity as it was observed in the experiments, see Figure~\ref{fig:cof}(b). It is also interesting a remark that the jump-off occurs in a regime where normal force weakly depends on probe-substrate distances. 

A relevant analytical model accounting for the behavior of the normal force with the probe-substrate distance has been elaborated by expressing the total surface energy of the droplet, $E$, as a function of its radius $r$, see Ref.~\cite{2014-Cheng,2015-Lee, ISRAELACHVILI2011415}. Then, $E\approx2\pi r h\gamma_{\rm w}-2\pi r^2\gamma_{\rm w}cos(\theta)$, where $\gamma_{\rm w}$ is the surface energy of water. In our model, the shape of the droplet is approximated as a cylinder with a volume $V=\pi r^2 h=340$~nm$^3$. Thus, the normal force considered as a function of the probe-substrate distance $h$ can be obtained by the straightforward derivative of the surface energy over $h$, $F_{N}(h)=-\partial E(h)/\partial h\approx\gamma_{\rm w} \sqrt{\pi V h}-2\gamma_{\rm w}\cos(\theta) V/h^2$.  Then, this analytical macroscopic model (represented by the dashed line) describes well the simulated data in Figure~\ref{fig:philicphobic}. The model takes as input the surface energy of water, given by the molecular dynamics SPC-model with $\gamma_{\rm m}=$52~mN/m~\cite{2007-Vega} and contact angles of 75$^{\rm o}$, 90$^{\rm o}$ and 105$^{\rm o}$. This model gives also insights into the dependence of the normal force on the volume of water in the contact and the probe-substrate distance. Then, by increasing the probe-substrate distance, the normal force changes its nature from repulsive into attractive interactions acting between the probe and the substrate. Below a probe-substrate distance noted $h_{0}=(4\cos^2(\theta)V)^{1/3}$, the surface tension acts to reduce the water surface in contact with the solid, as well as its overall surface. This tendency to obtain a more symmetric (spherical) shape results in a repulsive force between the probe and the substrate. In the example given configuration B1 in Figure~\ref{fig:philicphobic}(c), $h_0<$4~nm for a 340~nm$^3$ droplet. Above $h_0\approx$4~nm the droplet is increased and elongated, and its resistance against further extension results in an attractive (adhesive) force mediated by the water droplet between probe and substrate. The measured adhesive force in the experiment is the minimal (negative) normal force during the retraction of the probe. Our model shows that $F_{A}\propto -V^{1/3}$ at a probe to substrate distance $h_{\rm max}=4\cos^{2/3}(\theta)/\pi^{1/3}V^{1/3}$ (see Figure S5 in SI). The experimental adhesion force is independent of the sliding velocity (see Figure~\ref{fig:cof}), implying that the water volume of the droplet is also independent of the sliding velocity. From there, the model allows estimating the volume of the droplet in our experiments to be 8$\cdot$10$^6$~nm$^3$, and equivalent to a spherical droplet about $2r=$250~nm in diameter. Such a value is relevant considering that the experimental probe has a 0.2~\textmu m diameter and that the droplet may be not be completely spherical.

\section{Discussion}

Why water lubrication, {\it i.e.}, the modification of friction properties, is different on hydrophobic surfaces and hydrophilic surfaces? 
In the case of hydrophilic surfaces, the lubrication mechanism involving water is mainly related to pinning–depinning molecular processes at the substrate/capillary bridge contact line~\cite{2015-Kim,2015-Lee}. This mechanism yields a logarithmic increase of friction force with the sliding velocity. However, the origin of water lubrication on hydrophobic surfaces seems different. First, the morphology or shape of the capillary bridge resembles a cylinder with convex edges, as confirmed by the simulations (Figure~\ref{fig:philicphobic}). One also could not expect any binding of water molecules with the hydrophobic surface as the hydrogen bond between water molecules is stronger than the van der Waals water-substrate interaction. This assumption remains still valid even though the HOPG surface is commensurate with water yielding a certain amount of reorganization of water molecules within the contact~\cite{2016-Ye}. Eventually, both experimental and simulated observations showed that the principal mechanism describing the water-mediated friction in hydrophobic systems is related to the resistance to slip. Another interesting question is: what is the difference in the mechanism of water meniscus formation? Condensation was suggested as a mechanism responsible for the water meniscus growth~\cite{2002-Riedo} in hydrophilic surfaces. Our simulations, supported by the experimental results, indicate an alternative mechanism in which the water molecules adsorbed on the substrate accumulate inside the growing water capillary bridge during sliding. Simultaneously the experiment, there are other processes counterbalancing accumulation and removing water, such as evaporation.

\section{Conclusion}

In conclusion, the molecular dynamics simulations show that water molecules agglomerate in the hydrophobic nano-contact to form a droplet behaving as a liquid capillary bridge sliding on the substrate. The growth of this droplet is driven by collecting the adsorbed water on the substrate with the sliding AFM probe. The friction mechanism is related to a full slip regime of the droplet adsorbed on the AFM probe. Consequently, the friction force increases with the sliding velocity and the contact area between the droplet and substrate. In addition, the adhesion force does not depend on the sliding velocity. A simple analytic model considering the role of the water surface tension and the adhesive interactions with the solid surfaces highlights this behavior and allows for determining the volume of the water droplet in the contact. Furthermore, the model shows that the volume of the sliding droplet does not change with the sliding velocity. The experimental data are in good agreement with the simulations and this analytical model. \textcolor{black}{Eventually, our findings highlight valuable mechanisms which could lead to predictable dynamic friction involving hydrophobic contacts through control of the water volume within the contact by adjusting key parameters such as surface energy.}

\section{Methods}

\subsection{Circular mode atomic force microscope (CM-AFM)} 

The CM-AFM takes advantage of a circular motion of the AFM probe in the plane of the sample. If a relative vertical displacement of the probe with the substrate is imposed during the circular displacement, the friction law which represents the friction force {\it vs.} the normal applied load is instantaneously obtained with a high resolution. Simultaneously, one obtains the adhesion force which corresponds to the vertical interaction force between the probe and the sample surface as a jump-off of the probe is detected. Here, the CM-AFM was implemented on a Dimension 3100, Nanoscope V from Bruker AFM. With a constant circular motion frequency (100 Hz), the sliding velocity was varied by changing the diameter of the circular motion in the range of 18 nm to 3 \textmu m. The vertical displacement velocity of the probe, whose value, is 5~nm/s, is much lower than the sliding velocity. \textcolor{black}{Applying a vertical displacement to the probe with a scan velocity of 5~nm/s allows varying the normal load with a low velocity while the friction force is measured at a much larger sliding velocity (ranging from 10~\textmu m/s to 1~mm/s) and sliding displacement (ranging from 1~mm to 100~mm).} In such conditions, the friction force measurement for a given applied load is performed in a stationary regime without stop periods during the measurement and with a constant sliding velocity. The dependence of the adhesion and friction forces on the sliding velocity was investigated at room temperature with a constant relative humidity of 38\%. To this end, the probe was an AFM silicon nitride tip (DNP cantilevers from Bruker) with a radius $R$ of about 100~nm (according to scanning electronic microscopy (SEM) images). The normal spring constant, determined by the thermal noise method, was estimated to be 0.3~Nm$^{-1}$. \textcolor{black}{Backscatering SEM images} showed that the probe was rapidly covered by a contamination layer of carbon resulting from the interactions between the HOPG substrate and the nitride silicon probe. No other contamination or wear of the probe occurred after each experiment. \textcolor{black}{No other contamination or wear of the probe occurred after each experiment. This was checked by doing a force curve on a reference surface (clean silicon wafer) after each experiment for a given sliding velocity.}

\subsection{Sample preparation} 
The sample bulk HOPG material is in production annealed under pressure and high temperatures in production, removing all traces of water from within the structure which could influence its mechanical stability. The cleaving process takes off a thin layer of HOPG. This freshly cleaved surface has an atomically flat surface, with almost no structure steps visible, and provides a background with only carbon in the elemental signature. No plastic deformation of the sample was also observed. Samples of HOPG for the experiments were purchased from Fischer Scientific \textcolor{black}{and adequately cleaved before each experiment to avoid humidity hysteresis effect\cite{2018-Carpick}.} The local roughness of HOPG was determined through analysis of AFM topographic images to be 0.05~nm within a 5×5 \textmu m$^2$ area.

\subsection{Simulation} The whole atomistic model in this work hinges on the apex of the AFM tip above a hydrophobic plane. We assumed that the interaction with the probe is probably also hydrophobic due to contamination. Therefore, for simplicity, we have assumed that interactions between water/probe and water/substrate are the same. The intermolecular interactions between the water and solid phase (probe and substrate) are described {\it via} the Lennard-Jones potential. The different situations involving a contact angle with water of 75$^{\rm o}$, 90$^{\rm o}$ and 105$^{\rm o}$ are described by way of a  Lennard-Jones potential energy parameter respectively of $\epsilon=0.5$~kcal/mol, $\epsilon=0.45$~kcal/mol, and $\epsilon=0.35$~kcal/mol. The symmetry of the interactions regarding the probe with water and the substrate required a compatible atomic structure. The intra-molecular forces inside the water droplet are calculated with the SPC potential~\cite{2007-Vega}. An fcc crystalline confirmation for both the probe and a densely packed [111]-plane for the surface was chosen. The nearest-neighbor spacing of both the atoms of the substrate and probe is $d=2^{1/6}\sigma=4.01$~\AA. The relative positions of atoms in the substrate and the probe are fixed. The curvature of the probe is 20~nm. The in-plane size of the substrate is 20$\times$45~nm$^2$ for simulations dealing with the accumulation of water and 20$\times$20~nm$^2$ for an apt evaluation of the slip resistance of the water droplet and the adhesion forces. \textcolor{black}{The simulations of agglomeration of droplets under the probe were performed for a fixed probe-substrate distance. We created water droplets distributed on the surface by placing small groups of water molecules on the planar substrate and then running dynamics. The smaller water droplets moved and were attracted to one another and formed larger droplets; these simulations were executed to reach equilibrium, i.e. until the potential energy fluctuated around a constant value. For simulations in which we evaluated lateral and normal forces, harmonic springs in all three orthogonal directions connected the probe to the support, and the support moved at constant velocity along substrate, orthogonal to it, or simultaneously in two directions. The spring had stiffness of 2 N/m in all directions. The water contact angle was determined in separate simulations of a large droplet. The positions of the oxygen atoms were fitted around the perimeter of the contact with the surface, up to 8~\AA. The periodic boundary conditions are used in the substrate plane. The molecular dynamics (MD) simulations have been performed by way of the software package LAMMPS with time-steps of 2~fs with a Nose-Hoover thermostat set at 300~K~\cite{plimpton1995fast}.}

\begin{acknowledgement}

I.S. acknowledges support of Ministry of Education, Science, and
Technological Development of the Republic of Serbia through
the Institute of Physics Belgrade. Numerical calculations were run on the PARADOX super-computing facility at the Scientific Computing Laboratory of the Institute of Physics Belgrade. Authors are grateful to Dr. Robert Carpick for fruitful discussions and advice.

\end{acknowledgement}

\begin{suppinfo}

Complementary experimental and molecular dynamics data and characterization of the probe is provided.

\end{suppinfo}

\bibliography{achemso-demo}

\providecommand{\latin}[1]{#1}
\makeatletter
\providecommand{\doi}
  {\begingroup\let\do\@makeother\dospecials
  \catcode`\{=1 \catcode`\}=2 \doi@aux}
\providecommand{\doi@aux}[1]{\endgroup\texttt{#1}}
\makeatother
\providecommand*\mcitethebibliography{\thebibliography}
\csname @ifundefined\endcsname{endmcitethebibliography}
  {\let\endmcitethebibliography\endthebibliography}{}
\begin{mcitethebibliography}{36}
\providecommand*\natexlab[1]{#1}
\providecommand*\mciteSetBstSublistMode[1]{}
\providecommand*\mciteSetBstMaxWidthForm[2]{}
\providecommand*\mciteBstWouldAddEndPuncttrue
  {\def\EndOfBibitem{\unskip.}}
\providecommand*\mciteBstWouldAddEndPunctfalse
  {\let\EndOfBibitem\relax}
\providecommand*\mciteSetBstMidEndSepPunct[3]{}
\providecommand*\mciteSetBstSublistLabelBeginEnd[3]{}
\providecommand*\EndOfBibitem{}
\mciteSetBstSublistMode{f}
\mciteSetBstMaxWidthForm{subitem}{(\alph{mcitesubitemcount})}
\mciteSetBstSublistLabelBeginEnd
  {\mcitemaxwidthsubitemform\space}
  {\relax}
  {\relax}

\bibitem[Jinesh and Frenken(2008)Jinesh, and Frenken]{2008-Jinesh}
Jinesh,~K.~B.; Frenken,~J. W.~M. Experimental Evidence for Ice Formation at
  Room Temperature. \emph{Phys. Rev. Lett.} \textbf{2008}, \emph{101},
  036101\relax
\mciteBstWouldAddEndPuncttrue
\mciteSetBstMidEndSepPunct{\mcitedefaultmidpunct}
{\mcitedefaultendpunct}{\mcitedefaultseppunct}\relax
\EndOfBibitem
\bibitem[Ye \latin{et~al.}(2016)Ye, Egberts, Han, Johnson, Carpick, and
  Martini]{2016-Ye}
Ye,~Z.; Egberts,~P.; Han,~G.~H.; Johnson,~A. T.~C.; Carpick,~R.~W.; Martini,~A.
  Load-Dependent Friction Hysteresis on Graphene. \emph{ACS Nano}
  \textbf{2016}, \emph{10}, 5161--5168\relax
\mciteBstWouldAddEndPuncttrue
\mciteSetBstMidEndSepPunct{\mcitedefaultmidpunct}
{\mcitedefaultendpunct}{\mcitedefaultseppunct}\relax
\EndOfBibitem
\bibitem[Vilhena \latin{et~al.}(2016)Vilhena, Pimentel, Pedraz, Luo, Serena,
  Pina, Gnecco, and P{\'e}rez]{2016-Vilhena}
Vilhena,~J.~G.; Pimentel,~C.; Pedraz,~P.; Luo,~F.; Serena,~P.~A.; Pina,~C.~M.;
  Gnecco,~E.; P{\'e}rez,~R. Atomic-Scale Sliding Friction on Graphene in Water.
  \emph{ACS Nano} \textbf{2016}, \emph{10}, 4288--4293\relax
\mciteBstWouldAddEndPuncttrue
\mciteSetBstMidEndSepPunct{\mcitedefaultmidpunct}
{\mcitedefaultendpunct}{\mcitedefaultseppunct}\relax
\EndOfBibitem
\bibitem[Rhee \latin{et~al.}(2016)Rhee, Shin, and Jang]{2016-Rhee}
Rhee,~T.; Shin,~M.; Jang,~H. Effects of humidity and substrate hydrophilicity
  on nanoscale friction. \emph{Tribology International} \textbf{2016},
  \emph{94}, 234 -- 239\relax
\mciteBstWouldAddEndPuncttrue
\mciteSetBstMidEndSepPunct{\mcitedefaultmidpunct}
{\mcitedefaultendpunct}{\mcitedefaultseppunct}\relax
\EndOfBibitem
\bibitem[Arif \latin{et~al.}(2018)Arif, Colas, and Filleter]{2018-Arif}
Arif,~T.; Colas,~G.; Filleter,~T. Effect of Humidity and Water Intercalation on
  the Tribological Behavior of Graphene and Graphene Oxide. \emph{ACS Applied
  Materials {\&} Interfaces} \textbf{2018}, \emph{10}, 22537--22544\relax
\mciteBstWouldAddEndPuncttrue
\mciteSetBstMidEndSepPunct{\mcitedefaultmidpunct}
{\mcitedefaultendpunct}{\mcitedefaultseppunct}\relax
\EndOfBibitem
\bibitem[Hasz \latin{et~al.}(2018)Hasz, Ye, Martini, and Carpick]{2018-Carpick}
Hasz,~K.; Ye,~Z.; Martini,~A.; Carpick,~R.~W. Experiments and simulations of
  the humidity dependence of friction between nanoasperities and graphite: The
  role of interfacial contact quality. \emph{Phys. Rev. Materials}
  \textbf{2018}, \emph{2}, 126001\relax
\mciteBstWouldAddEndPuncttrue
\mciteSetBstMidEndSepPunct{\mcitedefaultmidpunct}
{\mcitedefaultendpunct}{\mcitedefaultseppunct}\relax
\EndOfBibitem
\bibitem[N'guessan \latin{et~al.}(2012)N'guessan, Leh, Cox, Bahadur, Tadmor,
  Patra, Vajtai, Ajayan, and Wasnik]{2012-Nguessan}
N'guessan,~H.~E.; Leh,~A.; Cox,~P.; Bahadur,~P.; Tadmor,~R.; Patra,~P.;
  Vajtai,~R.; Ajayan,~P.~M.; Wasnik,~P. Water tribology on graphene.
  \emph{Nature Communications} \textbf{2012}, \emph{3}, 1242\relax
\mciteBstWouldAddEndPuncttrue
\mciteSetBstMidEndSepPunct{\mcitedefaultmidpunct}
{\mcitedefaultendpunct}{\mcitedefaultseppunct}\relax
\EndOfBibitem
\bibitem[Pasumarty \latin{et~al.}(2011)Pasumarty, Johnson, Watson, and
  Adams]{2011-Pasumarty}
Pasumarty,~S.~M.; Johnson,~S.~A.; Watson,~S.~A.; Adams,~M.~J. Friction of the
  Human Finger Pad: Influence of Moisture, Occlusion and Velocity.
  \emph{Tribology Letters} \textbf{2011}, \emph{44}, 117\relax
\mciteBstWouldAddEndPuncttrue
\mciteSetBstMidEndSepPunct{\mcitedefaultmidpunct}
{\mcitedefaultendpunct}{\mcitedefaultseppunct}\relax
\EndOfBibitem
\bibitem[Li \latin{et~al.}(2013)Li, Wang, Kozbial, Shenoy, Zhou, McGinley,
  Ireland, Morganstein, Kunkel, Surwade, Li, and Liu]{Li2013}
Li,~Z.; Wang,~Y.; Kozbial,~A.; Shenoy,~G.; Zhou,~F.; McGinley,~R.; Ireland,~P.;
  Morganstein,~B.; Kunkel,~A.; Surwade,~S.~P.; Li,~L.; Liu,~H. Effect of
  airborne contaminants on the wettability of supported graphene and graphite.
  \emph{Nature Materials} \textbf{2013}, \emph{12}, 925--931\relax
\mciteBstWouldAddEndPuncttrue
\mciteSetBstMidEndSepPunct{\mcitedefaultmidpunct}
{\mcitedefaultendpunct}{\mcitedefaultseppunct}\relax
\EndOfBibitem
\bibitem[Vitorino \latin{et~al.}(2018)Vitorino, Vieira, Marques, and
  Rodrigues]{2018-Vitorino}
Vitorino,~M.~V.; Vieira,~A.; Marques,~C.~A.; Rodrigues,~M.~S. Direct
  measurement of the capillary condensation time of a water nanobridge.
  \emph{Scientific Reports} \textbf{2018}, \emph{8}, 13848\relax
\mciteBstWouldAddEndPuncttrue
\mciteSetBstMidEndSepPunct{\mcitedefaultmidpunct}
{\mcitedefaultendpunct}{\mcitedefaultseppunct}\relax
\EndOfBibitem
\bibitem[Szoszkiewicz and Riedo(2005)Szoszkiewicz, and Riedo]{2005-Riedo}
Szoszkiewicz,~R.; Riedo,~E. Nucleation Time of Nanoscale Water Bridges.
  \emph{Phys. Rev. Lett.} \textbf{2005}, \emph{95}, 135502\relax
\mciteBstWouldAddEndPuncttrue
\mciteSetBstMidEndSepPunct{\mcitedefaultmidpunct}
{\mcitedefaultendpunct}{\mcitedefaultseppunct}\relax
\EndOfBibitem
\bibitem[Feiler \latin{et~al.}(2007)Feiler, Stiernstedt, Theander, Jenkins, and
  Rutland]{2007-Feiler}
Feiler,~A.~A.; Stiernstedt,~J.; Theander,~K.; Jenkins,~P.; Rutland,~M.~W.
  Effect of Capillary Condensation on Friction Force and Adhesion.
  \emph{Langmuir} \textbf{2007}, \emph{23}, 517--522\relax
\mciteBstWouldAddEndPuncttrue
\mciteSetBstMidEndSepPunct{\mcitedefaultmidpunct}
{\mcitedefaultendpunct}{\mcitedefaultseppunct}\relax
\EndOfBibitem
\bibitem[Lee \latin{et~al.}(2015)Lee, Kim, Kim, and Jhe]{2015-Lee}
Lee,~M.; Kim,~B.; Kim,~J.; Jhe,~W. Noncontact friction via capillary shear
  interaction at nanoscale. \emph{Nature Communications} \textbf{2015},
  \emph{6}, 7359\relax
\mciteBstWouldAddEndPuncttrue
\mciteSetBstMidEndSepPunct{\mcitedefaultmidpunct}
{\mcitedefaultendpunct}{\mcitedefaultseppunct}\relax
\EndOfBibitem
\bibitem[Ho \latin{et~al.}(2011)Ho, Papavassiliou, Lee, and Striolo]{2011-Ho}
Ho,~T.~A.; Papavassiliou,~D.~V.; Lee,~L.~L.; Striolo,~A. Liquid water can slip
  on a hydrophilic surface. \emph{Proceedings of the National Academy of
  Sciences} \textbf{2011}, \emph{108}, 16170--16175\relax
\mciteBstWouldAddEndPuncttrue
\mciteSetBstMidEndSepPunct{\mcitedefaultmidpunct}
{\mcitedefaultendpunct}{\mcitedefaultseppunct}\relax
\EndOfBibitem
\bibitem[Thompson and Robbins(1989)Thompson, and Robbins]{1989-Thompson}
Thompson,~P.~A.; Robbins,~M.~O. Simulations of contact-line motion: Slip and
  the dynamic contact angle. \emph{Phys. Rev. Lett.} \textbf{1989}, \emph{63},
  766--769\relax
\mciteBstWouldAddEndPuncttrue
\mciteSetBstMidEndSepPunct{\mcitedefaultmidpunct}
{\mcitedefaultendpunct}{\mcitedefaultseppunct}\relax
\EndOfBibitem
\bibitem[Baudry \latin{et~al.}(2001)Baudry, Charlaix, Tonck, and
  Mazuyer]{2001-baudry}
Baudry,~J.; Charlaix,~E.; Tonck,~A.; Mazuyer,~D. Experimental Evidence for a
  Large Slip Effect at a Nonwetting Fluid-Solid Interface. \emph{Langmuir}
  \textbf{2001}, \emph{17}, 5232--5236\relax
\mciteBstWouldAddEndPuncttrue
\mciteSetBstMidEndSepPunct{\mcitedefaultmidpunct}
{\mcitedefaultendpunct}{\mcitedefaultseppunct}\relax
\EndOfBibitem
\bibitem[Teschke(2010)]{2010-Teschke}
Teschke,~O. Imaging Ice-Like Structures Formed on HOPG at Room Temperature.
  \emph{Langmuir} \textbf{2010}, \emph{26}, 16986--16990\relax
\mciteBstWouldAddEndPuncttrue
\mciteSetBstMidEndSepPunct{\mcitedefaultmidpunct}
{\mcitedefaultendpunct}{\mcitedefaultseppunct}\relax
\EndOfBibitem
\bibitem[Khan \latin{et~al.}(2010)Khan, Matei, Patil, and Hoffmann]{2010-Khan}
Khan,~S.~H.; Matei,~G.; Patil,~S.; Hoffmann,~P.~M. Dynamic Solidification in
  Nanoconfined Water Films. \emph{Phys. Rev. Lett.} \textbf{2010}, \emph{105},
  106101\relax
\mciteBstWouldAddEndPuncttrue
\mciteSetBstMidEndSepPunct{\mcitedefaultmidpunct}
{\mcitedefaultendpunct}{\mcitedefaultseppunct}\relax
\EndOfBibitem
\bibitem[Giovambattista \latin{et~al.}(2006)Giovambattista, Rossky, and
  Debenedetti]{2006-Giovambattista}
Giovambattista,~N.; Rossky,~P.~J.; Debenedetti,~P.~G. Effect of pressure on the
  phase behavior and structure of water confined between nanoscale hydrophobic
  and hydrophilic plates. \emph{Phys. Rev. E} \textbf{2006}, \emph{73},
  041604\relax
\mciteBstWouldAddEndPuncttrue
\mciteSetBstMidEndSepPunct{\mcitedefaultmidpunct}
{\mcitedefaultendpunct}{\mcitedefaultseppunct}\relax
\EndOfBibitem
\bibitem[Ouyang \latin{et~al.}(2018)Ouyang, de~Wijn, and Urbakh]{2018-Ouyang}
Ouyang,~W.; de~Wijn,~A.~S.; Urbakh,~M. Atomic-scale sliding friction on a
  contaminated surface. \emph{Nanoscale} \textbf{2018}, \emph{10},
  6375--6381\relax
\mciteBstWouldAddEndPuncttrue
\mciteSetBstMidEndSepPunct{\mcitedefaultmidpunct}
{\mcitedefaultendpunct}{\mcitedefaultseppunct}\relax
\EndOfBibitem
\bibitem[Shimizu \latin{et~al.}(2018)Shimizu, Maier, Verdaguer, Velasco-Velez,
  and Salmeron]{SHIMIZU201887}
Shimizu,~T.~K.; Maier,~S.; Verdaguer,~A.; Velasco-Velez,~J.-J.; Salmeron,~M.
  Water at surfaces and interfaces: From molecules to ice and bulk liquid.
  \emph{Progress in Surface Science} \textbf{2018}, \emph{93}, 87--107, Special
  Issue in Honor of Prof. Maki Kawai\relax
\mciteBstWouldAddEndPuncttrue
\mciteSetBstMidEndSepPunct{\mcitedefaultmidpunct}
{\mcitedefaultendpunct}{\mcitedefaultseppunct}\relax
\EndOfBibitem
\bibitem[Nasrallah \latin{et~al.}(2011)Nasrallah, Mazeran, and
  No\"el]{Noel-2011}
Nasrallah,~H.; Mazeran,~P.-E.; No\"el,~O. Circular mode: A new scanning probe
  microscopy method for investigating surface properties at constant and
  continuous scanning velocities. \emph{Review of Scientific Instruments}
  \textbf{2011}, \emph{82}, 113703\relax
\mciteBstWouldAddEndPuncttrue
\mciteSetBstMidEndSepPunct{\mcitedefaultmidpunct}
{\mcitedefaultendpunct}{\mcitedefaultseppunct}\relax
\EndOfBibitem
\bibitem[No\"el \latin{et~al.}(2012)No\"el, Mazeran, and Nasrallah]{2012-Noel}
No\"el,~O.; Mazeran,~P.-E.; Nasrallah,~H. Sliding Velocity Dependence of
  Adhesion in a Nanometer-Sized Contact. \emph{Phys. Rev. Lett.} \textbf{2012},
  \emph{108}, 015503\relax
\mciteBstWouldAddEndPuncttrue
\mciteSetBstMidEndSepPunct{\mcitedefaultmidpunct}
{\mcitedefaultendpunct}{\mcitedefaultseppunct}\relax
\EndOfBibitem
\bibitem[Eder \latin{et~al.}(2015)Eder, Feldbauer, Bianchi, Cihak-Bayr, Betz,
  and Vernes]{2015-Eder}
Eder,~S.~J.; Feldbauer,~G.; Bianchi,~D.; Cihak-Bayr,~U.; Betz,~G.; Vernes,~A.
  Applicability of Macroscopic Wear and Friction Laws on the Atomic Length
  Scale. \emph{Phys. Rev. Lett.} \textbf{2015}, \emph{115}, 025502\relax
\mciteBstWouldAddEndPuncttrue
\mciteSetBstMidEndSepPunct{\mcitedefaultmidpunct}
{\mcitedefaultendpunct}{\mcitedefaultseppunct}\relax
\EndOfBibitem
\bibitem[Ptak \latin{et~al.}(2019)Ptak, Almeida, and Prioli]{Ptak2019}
Ptak,~F.; Almeida,~C.~M.; Prioli,~R. Velocity-dependent friction enhances
  tribomechanical differences between monolayer and multilayer graphene.
  \emph{Scientific Reports} \textbf{2019}, \emph{9}, 14555\relax
\mciteBstWouldAddEndPuncttrue
\mciteSetBstMidEndSepPunct{\mcitedefaultmidpunct}
{\mcitedefaultendpunct}{\mcitedefaultseppunct}\relax
\EndOfBibitem
\bibitem[Green(1954)]{green}
Green,~M.~S. Markoff Random Processes and the Statistical Mechanics of
  Time‐Dependent Phenomena. II. Irreversible Processes in Fluids. \emph{The
  Journal of Chemical Physics} \textbf{1954}, \emph{22}, 398--413\relax
\mciteBstWouldAddEndPuncttrue
\mciteSetBstMidEndSepPunct{\mcitedefaultmidpunct}
{\mcitedefaultendpunct}{\mcitedefaultseppunct}\relax
\EndOfBibitem
\bibitem[Kubo(1957)]{kubo}
Kubo,~R. Statistical-Mechanical Theory of Irreversible Processes. I. General
  Theory and Simple Applications to Magnetic and Conduction Problems.
  \emph{Journal of the Physical Society of Japan} \textbf{1957}, \emph{12},
  570--586\relax
\mciteBstWouldAddEndPuncttrue
\mciteSetBstMidEndSepPunct{\mcitedefaultmidpunct}
{\mcitedefaultendpunct}{\mcitedefaultseppunct}\relax
\EndOfBibitem
\bibitem[Kim \latin{et~al.}(2015)Kim, Kwon, Lee, Kim, An, and Jhe]{2015-Kim}
Kim,~B.; Kwon,~S.; Lee,~M.; Kim,~Q.; An,~S.; Jhe,~W. Probing nonlinear rheology
  layer-by-layer in interfacial hydration water. \emph{Proceedings of the
  National Academy of Sciences} \textbf{2015}, \emph{112}, 15619--15623\relax
\mciteBstWouldAddEndPuncttrue
\mciteSetBstMidEndSepPunct{\mcitedefaultmidpunct}
{\mcitedefaultendpunct}{\mcitedefaultseppunct}\relax
\EndOfBibitem
\bibitem[Thompson and Troian(1997)Thompson, and Troian]{1997-Thompson}
Thompson,~P.~A.; Troian,~S.~M. A general boundary condition for liquid flow at
  solid surfaces. \emph{Nature} \textbf{1997}, \emph{389}, 360--362\relax
\mciteBstWouldAddEndPuncttrue
\mciteSetBstMidEndSepPunct{\mcitedefaultmidpunct}
{\mcitedefaultendpunct}{\mcitedefaultseppunct}\relax
\EndOfBibitem
\bibitem[Verdaguer \latin{et~al.}(2012)Verdaguer, Santos, Sauthier, Segura,
  Chiesa, and Fraxedas]{2012-Verdaguer}
Verdaguer,~A.; Santos,~S.; Sauthier,~G.; Segura,~J.~J.; Chiesa,~M.;
  Fraxedas,~J. Water-mediated height artifacts in dynamic atomic force
  microscopy. \emph{Phys. Chem. Chem. Phys.} \textbf{2012}, \emph{14},
  16080--16087\relax
\mciteBstWouldAddEndPuncttrue
\mciteSetBstMidEndSepPunct{\mcitedefaultmidpunct}
{\mcitedefaultendpunct}{\mcitedefaultseppunct}\relax
\EndOfBibitem
\bibitem[Cheng and Robbins(2014)Cheng, and Robbins]{2014-Cheng}
Cheng,~S.; Robbins,~M.~O. Capillary adhesion at the nanometer scale.
  \emph{Phys. Rev. E} \textbf{2014}, \emph{89}, 062402\relax
\mciteBstWouldAddEndPuncttrue
\mciteSetBstMidEndSepPunct{\mcitedefaultmidpunct}
{\mcitedefaultendpunct}{\mcitedefaultseppunct}\relax
\EndOfBibitem
\bibitem[Israelachvili(2011)]{ISRAELACHVILI2011415}
Israelachvili,~J.~N. \emph{Intermolecular and Surface Forces}, third edition
  ed.; Academic Press: Boston, 2011; pp 415--467\relax
\mciteBstWouldAddEndPuncttrue
\mciteSetBstMidEndSepPunct{\mcitedefaultmidpunct}
{\mcitedefaultendpunct}{\mcitedefaultseppunct}\relax
\EndOfBibitem
\bibitem[Vega and de~Miguel(2007)Vega, and de~Miguel]{2007-Vega}
Vega,~C.; de~Miguel,~E. Surface tension of the most popular models of water by
  using the test-area simulation method. \emph{The Journal of Chemical Physics}
  \textbf{2007}, \emph{126}, 154707\relax
\mciteBstWouldAddEndPuncttrue
\mciteSetBstMidEndSepPunct{\mcitedefaultmidpunct}
{\mcitedefaultendpunct}{\mcitedefaultseppunct}\relax
\EndOfBibitem
\bibitem[Riedo \latin{et~al.}(2002)Riedo, L\'evy, and Brune]{2002-Riedo}
Riedo,~E.; L\'evy,~F.; Brune,~H. Kinetics of Capillary Condensation in
  Nanoscopic Sliding Friction. \emph{Phys. Rev. Lett.} \textbf{2002},
  \emph{88}, 185505\relax
\mciteBstWouldAddEndPuncttrue
\mciteSetBstMidEndSepPunct{\mcitedefaultmidpunct}
{\mcitedefaultendpunct}{\mcitedefaultseppunct}\relax
\EndOfBibitem
\bibitem[Plimpton(1995)]{plimpton1995fast}
Plimpton,~S. Fast parallel algorithms for short-range molecular dynamics.
  \emph{Journal of computational physics} \textbf{1995}, \emph{117},
  1--19\relax
\mciteBstWouldAddEndPuncttrue
\mciteSetBstMidEndSepPunct{\mcitedefaultmidpunct}
{\mcitedefaultendpunct}{\mcitedefaultseppunct}\relax
\EndOfBibitem
\end{mcitethebibliography}

\end{document}